\documentclass[traditabstract,paper]{aa}
\usepackage{amsmath}
\usepackage{graphicx}
\usepackage{color}
\usepackage{natbib}

\bibpunct{(}{)}{;}{a}{}{,}

\begin{document}

\nocite{*}

\author{Edo van Uitert \inst{\ref{inst1},\ref{inst2}}\thanks{
        E-mail: vuitert@ucl.ac.uk} and Peter Schneider \inst{\ref{inst2}}}
\institute{University College London, Gower Street, London WC1E 6BT, UK \label{inst1} \and Argelander-Institut f\"ur Astronomie, Auf dem H\"ugel 71, 53121 Bonn, Germany \label{inst2}
}

\title{Systematic tests for position-dependent additive shear bias}

\abstract{We present new tests to identify stationary position-dependent additive shear biases in weak gravitational lensing data sets. These tests are important diagnostics for currently ongoing and planned cosmic shear surveys, as such biases induce coherent shear patterns that can mimic and potentially bias the cosmic shear signal. The central idea of these tests is to determine the average ellipticity of all galaxies with shape measurements in a grid in the pixel plane. The distribution of the absolute values of these averaged ellipticities can be compared to randomised catalogues; a difference points to systematics in the data. In addition, we introduce a method to quantify the spatial correlation of the additive bias, which suppresses the contribution from cosmic shear and therefore eases the identification of a position-dependent additive shear bias in the data. We apply these tests to the publicly available shear catalogues from the Canada-France-Hawaii Telescope Lensing Survey (CFHTLenS) and the Kilo Degree Survey (KiDS) and find evidence for a small but non-negligible residual additive bias at small scales. As this residual bias is smaller than the error on the shear correlation signal at those scales, it is highly unlikely that it causes a significant bias in the published cosmic shear results of CFHTLenS. In CFHTLenS, the amplitude of this systematic signal is consistent with zero in fields where the number of stars used to model the point spread function (PSF) is higher than average, suggesting that the position-dependent additive shear bias originates from undersampled PSF variations across the image.
}

\titlerunning{Systematic tests for position-dependent $c$}

\maketitle

\section{Introduction}
\indent Cosmic shear, the coherent distortion of the observed shapes of distant galaxies by the gravitational field of intervening matter distributions, is one of the most powerful tools to constrain cosmological parameters \citep{Albrecht06}. Since its first detection in the early 2000s, the field has rapidly expanded and matured \citep[see][for a recent review]{Kilbinger15}. Several large and deep optical imaging surveys are currently ongoing \citep[KiDS, DES and HSC, see][respectively]{Kuijken15,Jarvis15,Miyazaki15} or will begin in the near future \citep[e.g. Euclid, LSST and WFIRST, see][respectively]{Laureijs11,LSST09,Spergel15} that are designed to measure cosmic shear. These surveys will map large portions of sky to great depths, increasing the number of galaxies usable for weak lensing by a factor of 100-1000 compared to the current state-of-the-art in cosmic shear: the Canada-France-Hawaii Lensing Survey \citep[CFHTLenS;][]{Heymans12}, an analysis of 139 deg$^2$ of high-quality optical imaging data, which yielded the tightest lensing constraints on cosmological parameters to date \citep{Kilbinger13,Heymans13,Kitching14}. Consequently, the precision of the shear measurements in these new surveys will be pushed down by orders of magnitude. Demonstrating that systematic errors in the lensing measurements are under control is imperative before the data can be exploited for cosmology. \\
\indent The accuracy of weak-lensing shape measurements is commonly quantified with two numbers, the multiplicative bias $m$ and the additive bias $c$, following \mbox{$ \langle \epsilon_{\rm g} \rangle = (1+m)\times \gamma + c$} \citep{Heymans06}. In an unbiased shape measurement method, $m=c=0$, such that the observed galaxy ellipticities $\epsilon_{\rm g}$ form an unbiased estimate of the shear $\gamma$. No shear measurement method to date has proven to be unbiased when tested under realistic conditions. Even worse, $m$ and $c$ are generally not constant, but depend on the flux of a galaxy, its size, the ellipticity of the point spread function (PSF), the Strehl-ratio, the sky background, etc. The dependencies on these parameters need to be extremely well calibrated to enable an unbiased cosmological exploitation of the data. \\
\indent There are several causes for additive and multiplicative shear biases \citep[see e.g.][]{Massey13}. Multiplicative biases are mainly caused by noise: measuring an ellipticity generally involves a non-linear transformation of pixel data, a process in which pixel noise enters non-linearly and does not average out \citep{Melchior12,Refregier12}. Another source of multiplicative bias is model bias: the implicit or explicit use of incorrect galaxy shape models to fit to the observed galaxies \citep{Bernstein10}. Additive biases can be caused by elliptical PSFs or by charge transfer inefficiencies that have not been completely accounted for. The magnitude of these biases generally depends on the shape measurement method that is employed. How these shape measurement biases propagate in cosmic shear studies is investigated in several works \citep[see e.g.][and references therein]{Massey13,Cropper13,Kitching16}. \\
\indent There are different strategies to determine $m$ and $c$: $m$ is usually determined by applying the shape measurement method to simulated data that mimics the real data as closely as possible. By comparing the recovered shear for different known input shear values, $m$ can be determined. Alternatively, cross-correlations of the lensing maps with galaxy density and cosmic microwave background lensing maps can be used \citep{Das13,Liu16}, as well as cross-correlations with weak lensing magnification \citep{Rozo10,Vallinotto10} or generalised shear-ratio tests \citep{Schneider16}. Additive shear bias can be determined from the data itself. Since galaxies do not have a preferred direction on the sky, their mean ellipticity should average to zero. Hence by determining the mean galaxy ellipticity, additive biases can be identified and subsequently removed by subtracting it from the observed galaxy ellipticities. \\
\indent Additive biases are usually determined as a function of galaxy property and observing condition, but not as a function of position on the camera. PSF ellipticities and star densities usually vary in the field-of-view and could cause a position-dependent additive bias that is stationary between exposures. Previously employed correction schemes that only measured a field-averaged additive bias would miss position-dependent residuals and these could still be present in the data. A coherent additive bias pattern is problematic as it might mimic the real, physical correlation between galaxy shapes due to cosmic shear, and could therefore bias cosmic shear analyses \citep{Kitching16}. Motivated by this concern, we develop tests to identify whether a position-dependent additive shear bias (which we refer to as position-dependent $c$ from here on) is present in weak lensing data sets. \\
\indent In Sect. \ref{sec_meth} we present a method to identify position-dependent $c$. We apply it to data from CFHTLenS in Sect. \ref{sec_1d}. In Sect. \ref{sec_2d}, we study the spatial correlation of the additive bias and investigate its dependence on position in the field, stellar density and photometric redshift. We repeat our analysis on KiDS data in Sect. \ref{sec_kids} and conclude in Sect. \ref{sec_concl}. We assume that the reader is familiar with the basics of gravitational lensing. For a general introduction, please see \citet{Bartelmann01}.

\section{Methodology \label{sec_meth}}
The central idea of our tests is to determine the average ellipticities of galaxies in a pixel grid on the detector and analyse their properties. If a position-dependent $c$ is present that is stationary between exposures (e.g. in one corner of the image where the PSF ellipticity is always large), this both affects the distribution of absolute values of the averaged ellipticities, and also causes a positive correlation between the average ellipticities of neighbouring grid cells at small separations.  \\
\indent We started with defining a regularly spaced grid in the pixel plane. For a single image (i.e. a pointing on the sky) called S, we determined the average ellipticity of all galaxies in each grid cell:
\begin{equation}
\langle \epsilon^{\rm S}\rangle({\bf x}_i)=\frac{\sum_{k \in i} \epsilon_{k}w_{k}}{\sum_{k \in i}w_{k}},
\end{equation}
with ${\bf x}_i$ the position of grid cell $i$, $\epsilon_{k}$ the complex ellipticity and $w_k$ the lensing weight of galaxy $k$. The sums run over all galaxies $k$ in image S that fall inside grid cell $i$. \\
\indent Next, we determined the mean ellipticity in each grid cell averaged over all images:
\begin{equation}
\langle E \rangle({\bf x}_i)=\frac{\sum_{s}\sum_{k \in i} \epsilon_{k}w_{k}}{\sum_{s}\sum_{k \in i}w_{k}},
\end{equation}
which explicitly sums over all images $s$. We also defined the average ellipticity of all images except one, image S:
\begin{equation}
\langle  E^{\rm notS} \rangle ({\bf x}_i)=\frac{\sum_{s, s \ne {\rm S }}\sum_{k \in i} \epsilon_{k}w_{k}}{\sum_{s, s \ne {\rm S}}\sum_{k \in i}w_{k}}.
\end{equation}
Using these average ellipticities, we defined two systematic shear correlation functions:
\begin{equation}
\xi_{\rm sys}^{{\rm tt}, {\rm S}}(\theta)=\frac{\sum_{i} \sum_{j} {\sf w}_i {\sf w}_j \langle \epsilon_{\rm t}^{\rm S}\rangle ({\bf x}_i) \langle E_{\rm t}^{\rm notS}\rangle ({\bf x}_j)}{\sum_{i} \sum_{j} {\sf w}_i {\sf w}_j},
\end{equation}
and
\begin{equation}
\xi_{\rm sys}^{\times\times, {\rm S}}(\theta)=\frac{\sum_{i} \sum_{j} {\sf w}_i {\sf w}_j \langle \epsilon_\times^{\rm S}\rangle ({\bf x}_i) \langle E_\times^{\rm notS}\rangle ({\bf x}_j)}{\sum_{i} \sum_{j} {\sf w}_i {\sf w}_j},
\end{equation}
with $i$ and $j$ denoting different grid cells, $\langle \epsilon_{\rm t}^{\rm S}\rangle$ and $\langle \epsilon_\times^{\rm S}\rangle$ the tangential and cross component of $\langle \epsilon^{\rm S}\rangle$ measured relative to the separation vector between ${\bf x}_i$ and ${\bf x}_j$, and $\langle E_{\rm t}^{\rm notS}\rangle$ and $\langle E_\times^{\rm notS}\rangle$ the tangential and cross component of $\langle E^{\rm notS}\rangle$. For the definition of tangential and cross shear, please see for example \citet{Kilbinger15}. $\theta$ is the pixel separation between grid cell $i$ and $j$. ${\sf w}_i$ and ${\sf w}_j$ are weight factors, which equal the sum of the weights $w_{k}$ of galaxies that went into computing $\langle \epsilon_{{\rm t}/\times}^{\rm S}\rangle({\bf x}_i)$ and $\langle  E_{{\rm t}/\times}^{\rm notS} \rangle ({\bf x}_j)$, respectively. The sum runs over the grid cells whose separation falls inside a $\theta$ bin. From these we formed
\begin{equation}
\xi_{\rm sys}^{+/-, {\rm S}}=\xi_{\rm sys}^{{\rm tt}, {\rm S}}\pm\xi_{\rm sys}^{\times\times, {\rm S}},
 \label{eq_xisys}
\end{equation}
which is our estimator for position-dependent $c$ for a single image S. We note that $\xi_{\rm sys}^{+, {\rm S}}$ is equal to the correlation of the two grid-averaged ellipticities, $\xi_{\rm sys}^{+, {\rm S}}=\langle \Re \left[ \epsilon^{\rm S}E^{\rm notS*}\right] \rangle=$ \mbox{ $ \langle \epsilon_1^{\rm S}\times E_1^{\rm notS} +  \epsilon_2^{\rm S}\times E_2^{\rm notS} \rangle$}. \\
\indent We also measured the complex part of the systematic shear correlation function:
\begin{equation}
\xi_{\rm sys}^{{\rm t}\times, {\rm S}}(\theta)=\frac{\sum_{i} \sum_{j} {\sf w}_i {\sf w}_j \langle \epsilon_{\rm t}^{\rm S}\rangle ({\bf x}_i) \langle E_\times^{\rm notS}\rangle ({\bf x}_j)}{\sum_{i} \sum_{j} {\sf w}_i {\sf w}_j},
\end{equation}
and
\begin{equation}
\xi_{\rm sys}^{\times {\rm t}, {\rm S}}(\theta)=\frac{\sum_{i} \sum_{j} {\sf w}_i {\sf w}_j \langle \epsilon_\times^{\rm S}\rangle ({\bf x}_i) \langle E_{\rm t}^{\rm notS}\rangle ({\bf x}_j)}{\sum_{i} \sum_{j} {\sf w}_i {\sf w}_j}.
\end{equation}
Their equivalents in cosmic shear studies are usually not measured, since they are expected to vanish from parity symmetry. In the presence of systematics, that is no longer necessarily the case. \\
\indent To estimate a survey-averaged systematic shear signal, $\xi_{\rm sys}^{+/-}$, we used a bootstrap technique. For a survey of $N$ images, we randomly drew $N$ systematic correlation functions ($\xi_{\rm sys}^{+/-, {\rm S}}$) from the full set with replacement. For each bootstrap realisation we determined the average systematic correlation function, $\xi_{\rm sys}^{+/-}$. In total, we created $10\,000$ bootstrap realisations. Their mean is our systematic signal, the scatter between the bootstrap realisations forms the error. $\xi_{\rm sys}^{{\rm t}\times/\times {\rm t}}$ was determined in a similar fashion.  \\
\indent The reason why we correlate the average ellipticity of galaxies in a single image to the average ellipticity of all other images, is that it suppresses the contribution from cosmic shear. Had we instead correlated the average ellipticities of all images, the galaxies in neighbouring grid cells from the same image would be subject to the same cosmic shear field, `contaminating' $\xi_{\rm sys}^{+/-}$ with a real, cosmic shear signal. We demonstrate  in Sect. \ref{sec_csrem} that this effect is small but not entirely negligible. The remaining cosmic shear contribution to our estimator, which is only present at scales larger than the size of an image, is even smaller and can be safely ignored. In principle, it could be further suppressed by excluding not only image S in $E_1^{\rm notS}$, but also its neighbours. \\
\indent We note that alternative estimators of the systematic correlation functions could be formed as well; one could, for example, correlate the average ellipticities of pairs of images S and T (for S$\neq$T), and randomly draw from those correlation functions to form a survey average. Alternatively, one could determine the average ellipticities of two images, and correlate that to the average ellipticities of the other images. This effectively boils down to adjusting the weighting scheme, which we plan to explore in a future work.

\section{1-point statistics \label{sec_1d}}
\begin{figure*}
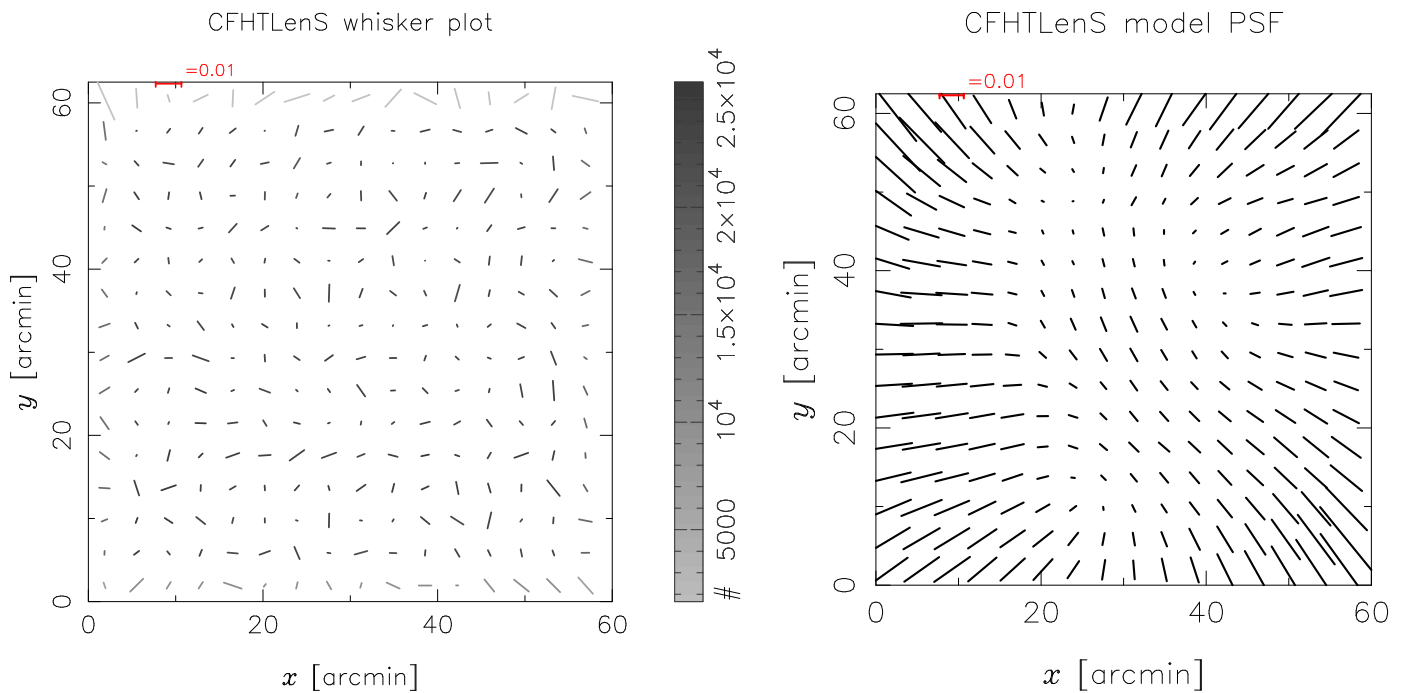

  \begin{minipage}[t]{0.49\linewidth}
   \includegraphics[width=\linewidth,angle=-90]{plot_CFHTLenS_whisk_16.ps}
  \end{minipage}
  \hspace{10mm}
  \begin{minipage}[t]{0.49\linewidth}
   \includegraphics[width=\linewidth,angle=-90]{plot_CFHTLenS_whisk_16_PSF.ps}
  \end{minipage}
   \caption{{\it Left-hand panel}: whisker plot showing the average galaxy ellipticity as a function of field position in the 128 CFHTLenS `pass' fields for the G2 grid (16$\times$16). The sticks indicate the size and orientation of the averaged ellipticities. The grey-scale of the sticks indicate the number of galaxies in a grid cell. The range of the horizontal and vertical axis corresponds to the size of a CFHTLenS image. {\it Right-hand panel}: whisker plot of the average model PSF in CFHTLenS obtained by averaging the model PSFs at the location of the galaxies for the same grid. }
   \label{plot_whisker}
\end{figure*}
\begin{figure*}
   \includegraphics[width=0.45\linewidth,angle=-90]{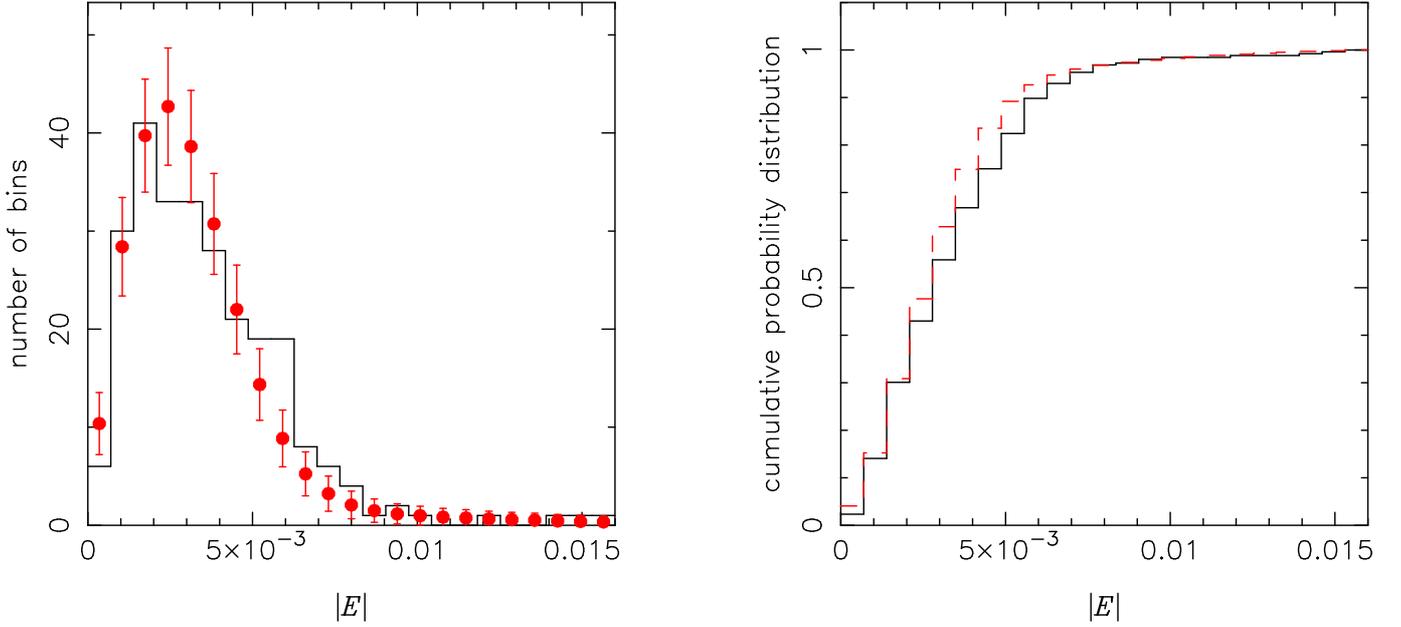}
   \caption{{\it Left-hand panel}: Histogram of mean ellipticity values in the G2 grid. The black line shows the distribution for the original CFHTLenS catalogue, the red dots indicate the mean and scatter of $10\,000$ realisations where the ellipticities were randomised. {\it Right-hand panel}: cumulative probability distribution of the observed and randomised histograms.}
   \label{plot_ellhist}
\end{figure*}

We tested our estimator on the publicly available shape measurements catalogues from CFHTLenS \citep{Heymans12}. The catalogues are based on 139 deg$^2$ of imaging data in the $ugrzi$-bands from the CFHT Legacy Survey, obtained with MegaPrime, a multi-chip camera that consists of 9$\times$4 CCDs of 2048$\times$4096 pixels each with a pixel scale of 0.187 arcsec. The total field-of-view is roughly 1 deg$^2$. The lensing measurements were performed on the $i$-band data, using a shape measurement method called \emph{lens}fit \citep{Miller13}, a Bayesian forward-modelling technique which models galaxies as a bulge plus disc, applies a shear and convolves them with a model PSF that is determined from the stars in the images \citep{Miller07,Kitching08}. The inverse variance weights provided by \emph{lens}fit are applied when we compute the mean ellipticities. The five-band photometry was used to derive photometric redshifts of galaxies with the BPZ method \citep{Benitez00}. The photometric redshifts were found to be robust in the range $0.2<z_{\rm B}<1.3$ \citep{Hildebrandt12}, with $z_{\rm B}$ the peak of the posterior redshift distribution. We use all galaxies with a non-zero \emph{lens}fit weight and with a photometric redshift $0.2<z_{\rm B}<1.3$; the effective weighted galaxy number density is 11 arcmin$^{-2}$. We limit ourselves to the 128 fields that passed the systematic test of \citet{Heymans12}; the 43 fields that did not pass this test are analysed separately in Sec. \ref{sec_fail}. \\
\indent The shape measurement catalogues from CFHTLenS have a non-zero multiplicative and additive shear bias. The multiplicative bias is determined by applying \emph{lens}fit to image simulations that mimic the actual observations. We ignore it here as we only focus on the additive bias. The additive bias in CFHTLenS is negligible in $\epsilon_1$ \mbox{($c_1=0.0001\pm0.0001$)} but not in $\epsilon_2$, where it has a value of $c_2=0.0020\pm0.0001$. This bias is found to scale with galaxy size $r$ and signal-to-noise $\nu_{\rm SN}$, but not with PSF size, PSF ellipticity and galaxy type. \citet{Heymans12} fit a functional form to model the dependence on $r$ and $\nu_{\rm SN}$:
\begin{equation}
c_2=\max \left[ \frac{11.910\log_{10}(\nu_{\rm SN})-12.715}{1+\left(\frac{r}{0.01''}\right)^{2.458}} , 0\right].
\end{equation}
The bias is predicted on a galaxy-by-galaxy basis and is provided as a separate column in the public catalogue. We corrected the ellipticities with this correction factor by subtracting it before we started our tests. \\
\indent We defined three regularly spaced grids on each field that contain 8$\times$8, 16$\times$16 and 32$\times$32 grid cells. In what follows, we refer to these as the G1, G2 and G3 grids, respectively. These grids were designed to enable us to roughly trace the chip gaps. To define the grid, we used the minimum and maximum $x$- and $y$-positions of all galaxies that passed our selection criteria. As the $i$-band data consists of seven dithered exposures on average, with a dithering step that is larger in the vertical direction (up to $\sim$3 arcmin) than in the horizontal direction (up to $\sim$0.5 arcmin) to fill in the larger gap between the chip rows, our grid is slightly rectangular. The horizontal axis spans a range of 58.8 arcmin, whilst the vertical axis spans a range of 62.5 arcmin. \\
\indent The whisker plot of $\langle E \rangle({\bf x}_i)$ for the G2 grid is shown in Fig. \ref{plot_whisker}. The bins at the edges contain roughly half the number of galaxies compared to the central ones and therefore have a larger scatter. In the absence of a position-dependent $c$, there should be no pattern in this plot. By eye, we can identify a number of suspicious features, such as at the bottom right-hand corner, where several grid cells have a similar $\langle E_2 \rangle$ component, and at $y$-values just below 20 arcmin, where we observe a row of grid cells with similar $\langle E_1 \rangle$ values. \\
\indent To check whether these features are significant, we first quantified the distribution of the absolute values of the average ellipticities in the grid cells, \mbox{$|E| \equiv |\langle E \rangle | =\sqrt{\langle E_1 \rangle^2+ \langle E_2 \rangle^2}$}. We show the histogram of $|E|$ in the left-hand panel of Fig. \ref{plot_ellhist}. To see if they are suspiciously large, we randomised the ellipticities: every galaxy in our sample was assigned the ellipticity and weight of another galaxy that was randomly drawn from the full CFHTLenS catalogue. We remeasured $|E|$ using the same grid and determined the histogram. The advantage of this procedure is that it uses the true observed ellipticity distribution of CFHTLenS, and that the density of galaxies in each bin is preserved. We repeated this procedure $10\,000$ times. The mean of these randomised histograms is shown with red circles in the left-hand panel of Fig. \ref{plot_ellhist}, the error bars indicate the scatter.  \\
\indent The histogram of $|E|$ for the original CFHTLenS catalogues differs from the mean of the randomised ones. We find fewer bins with $|E|<4\times10^{-3}$ in the original histogram and more bins with $|E|>5\times10^{-3}$. If a position-dependent $c$ is present, these tend to increase the average ellipticity of galaxies in certain regions and one expects an increase of bins with large mean ellipticity and a corresponding decrease of bins with small ellipticities. In the random catalogues, any position-dependent $c$ is averaged out. Hence our results are indicative of a position-dependent $c$. \\
\indent To quantify the difference between the histograms, we computed the reduced $\chi^2$ between the observed and the mean of the randomised histograms, using the scatter between the random realisations as errors. We find that $\chi^2_{\rm red}=1.44$ (with 23 degrees of freedom). The corresponding probability to exceed ($p$-value) is 0.08, which provides weak evidence that the observed distribution is not a random realisation. \\
\indent The $\chi^2_{\rm red}$ test is ignorant about the sign of the difference between the two samples (a systematic decrease of the number of bins at $|E|<4\times10^{-3}$ and a systematic excess at $|E|>5\times10^{-3}$). A more sensitive test is the Kolmogorov-Smirnov (KS) statistic on two samples. We determined the cumulative probability of the two histograms and show it in the right-hand panel of Fig. \ref{plot_ellhist}. We measured the KS statistic, which is the maximum distance between the cumulative probability distributions, to check whether the two distributions are drawn from the same reference distribution. The KS statistic has a value of 0.097, with a corresponding $p$-value of 0.015, which shows that the two distributions are different. \\
\indent If galaxies at the edge of the field are systematically noisier and more elliptical, we may overestimate how odd/unlikely the observed distribution is. We therefore also made random catalogues by only rotating the galaxy ellipticities with random amounts (but keeping the magnitude of the ellipticity fixed to the input value). We made $10\,000$ random catalogues and assessed the difference between the original distribution and the mean of the random distributions as before. In this case, the reduced $\chi^2$ between the observed and the mean of the randomised histograms has a value of 1.39 and the corresponding KS statistic has a value of 0.096. Our results therefore do not depend on how we create the random catalogues.

\section{2-point statistics \label{sec_2d}}
\begin{figure*}
  \includegraphics[width=0.95\linewidth]{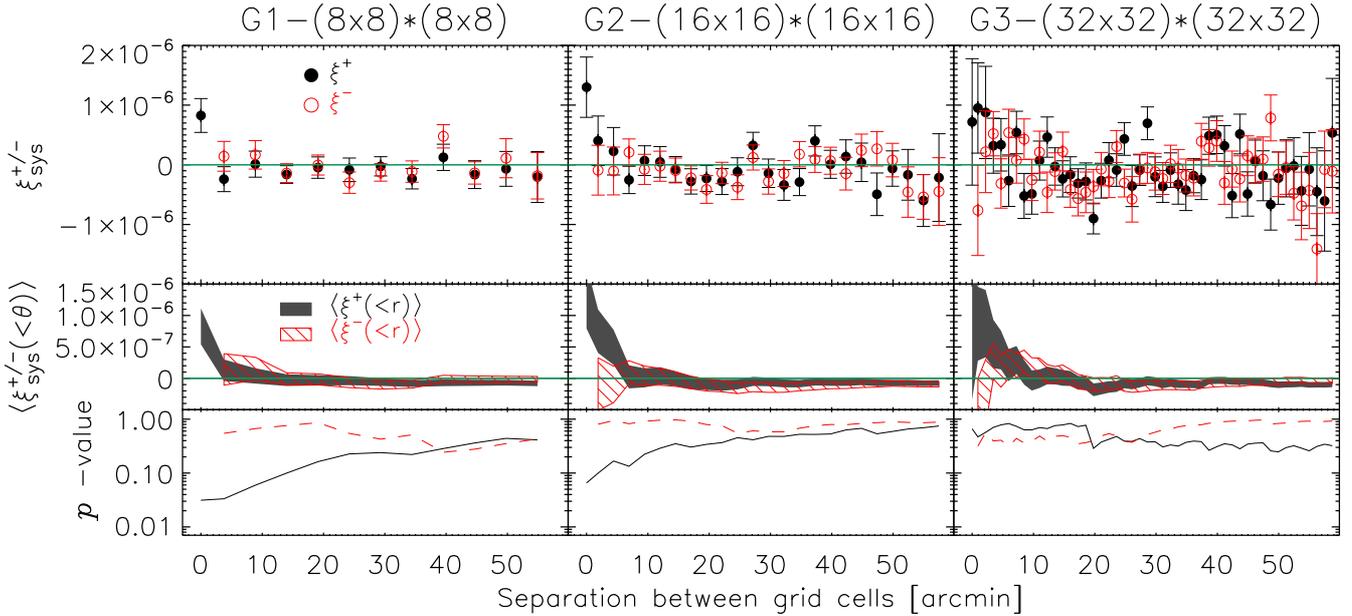}
  \caption{$\xi_{\rm sys}^{+/-}$ correlation function as a function of separation between grid cells. A non-zero signal indicates the presence of a position-dependent $c$ with power at that scale. The left-hand, middle and right-hand panels show the signal for the G1, G2 and G3 grids, respectively. The x-axis range corresponds to the size of the image in all three panels, the first radial bin shows $\xi_{\rm sys}^{+}$  at zero lag. The errors indicate the scatter between bootstrap realisations. The middle row shows the weighted mean of $\xi_{\rm sys}^{+/-}$ and its 68\% confidence intervals, determined using all radial bins up to the one of interest. The bottom row shows the $p$-values of the null hypothesis, with the solid black (red-dashed) line for $\langle \xi_{\rm sys}^{+}(<r)\rangle$ ($\langle\xi_{\rm sys}^{-}(<r)\rangle$).}
  \label{plot_ecorr}
\end{figure*}
\begin{figure*}
  \includegraphics[width=0.95\linewidth]{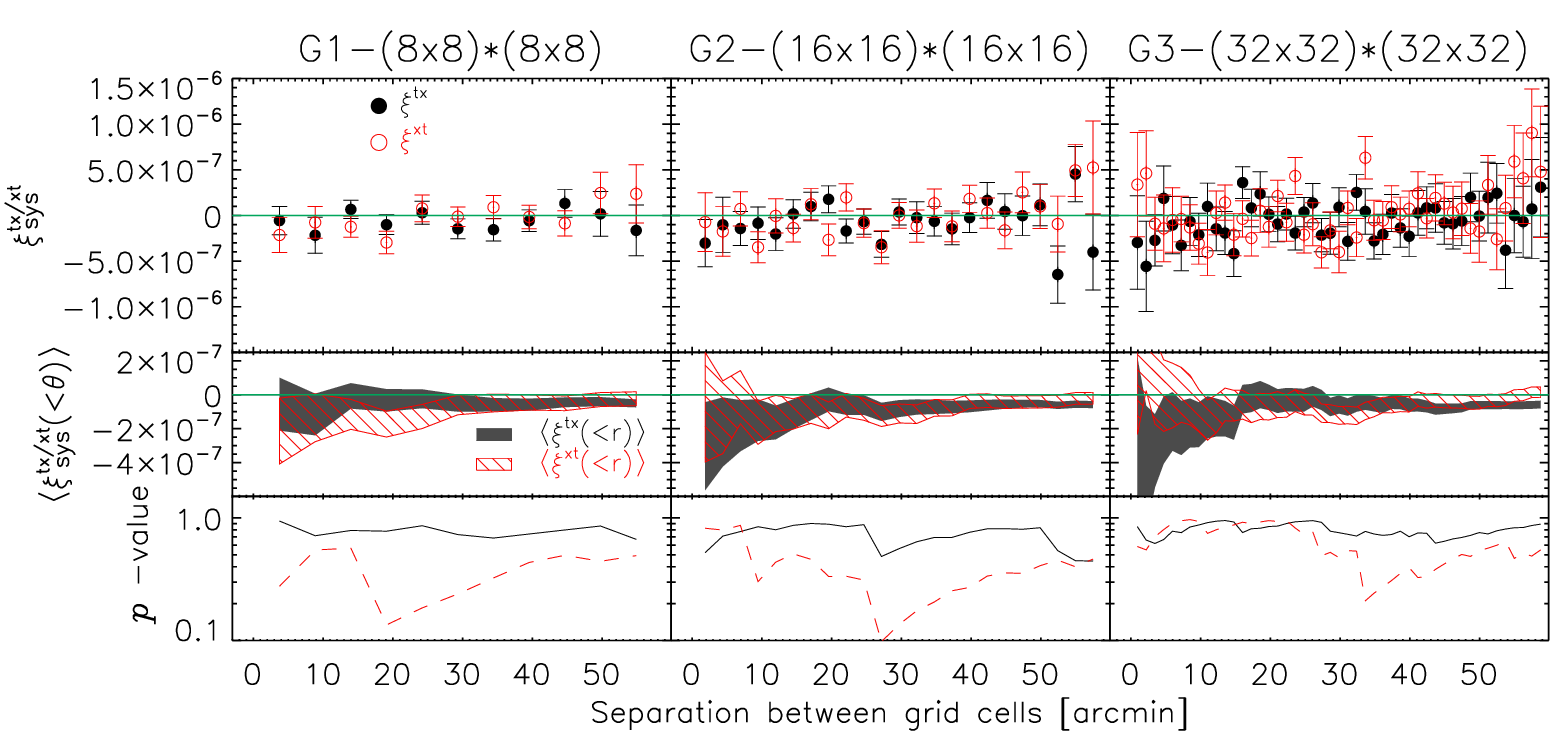}
  \caption{$\xi_{\rm sys}^{{\rm t}\times/\times {\rm t}}$ correlation function as a function of separation between grid cells. A non-zero signal indicates the presence of a position-dependent $c$ with power at that scale. The left-hand, middle and right-hand panels show the signal for the G1, G2 and G3 grids, respectively. The x-axis range corresponds to the size of the image in all three panels. The errors indicate the scatter between bootstrap realisations. The middle row shows the weighted mean of $\xi_{\rm sys}^{{\rm t}\times/\times {\rm t}}$ and its 68\% confidence intervals, determined using all radial bins up to the one of interest. The bottom row shows the $p$-values of the null hypothesis, with the solid black (red-dashed) line for $\langle \xi_{\rm sys}^{{\rm t}\times}(<r)\rangle$ ($\langle\xi_{\rm sys}^{\times {\rm t}}(<r)\rangle$).}
  \label{plot_ecorrIM}
\end{figure*}
\indent We quantified the presence of position-dependent $c$ using the systematic correlation functions from Eq. (\ref{eq_xisys}) as a function of separation between the grid cells. We used 12, 24 and 48 radial bins between $(0,60]$ arcmin for the G1, G2 and G3 grids, respectively. The first bin contains $\xi_{\rm sys}^{+}$ at zero lag. $\xi_{\rm sys}^{-}$ is not defined at zero lag and hence not shown. The remaining radial bins are evenly spaced up to 60 arcmin. The correlation functions are shown in Fig. \ref{plot_ecorr}, where the three columns correspond to the three grids. The middle panel shows the value of a constant fitted to $\xi_{\rm sys}^{+/-}$, using scales up to that radius (e.g. the third radial bin includes the measurements of the first, second and third radial bins; the final point is the weighted mean of $\xi_{\rm sys}^{+/-}$ averaged over the entire range). The absence of a position-dependent $c$ would result in a constant that is consistent with zero on all scales.  When we fit the constant we use the covariance matrix of $\xi_{\rm sys}^{+/-}$, determined from the bootstrap realisations. $\xi_{\rm sys}^{+}$ is weakly correlated between neighbouring radial bins and $\xi_{\rm sys}^{-}$ is practically uncorrelated. We corrected the inverse of the covariance matrix with a correction factor that accounts for a bias in the inversion that arises from noise \citep{Kaufmann67,Hartlap07}. Finally, in the third row we show the $p$-values that correspond to the $\chi^2$ of the null hypothesis. A $p$-value that is much smaller than 1 indicates that the data is not consistent with zero. Here and in the following, we calculate the $p$-values using a calibration described in Appendix \ref{app_p}. \\
\indent $\xi_{\rm sys}^{+}$ is significantly non-zero at small separations. The first bin, which shows the correlation at zero lag, deviates from zero with 2.9$\sigma$, 2.6$\sigma$ and 0.7$\sigma$ for the G1, G2 and G3 grids, respectively. This shows that most of the power of the small-scale additive bias has a scale-length that corresponds to the size of the G1 grid-cell, $60/8=7.5$ arcmin, roughly the width of a chip. Our results therefore point to an additive bias that is constant per chip, but varies between chips. This suggests that the bias originates from the constant term in the PSF modelling that is fit per chip (see also Sec. \ref{sec_star}). \\
\indent Refining the grid reveals more features which are smoothed out in the more crudely sampled grids. Furthermore, the finer grids appear to reveal the presence of additional structure, most noticeably a negative dip at a radial separation of $\sim$20 arcmin. Interestingly, $\sim$20 arcmin is not obviously related to a structure of the camera such as the size of a chip. \\
\indent In contrast to $\xi_{\rm sys}^{+}$, $\xi_{\rm sys}^{-}$ depends on the direction of the separation vector between the grid cells. We therefore might expect to see some differences between the trends in the three grids. In all cases, $\xi_{\rm sys}^{-}$ does not show an obvious trend; the incremental weighted mean is consistent with zero when averaged over the full radial range. \\
\indent The complex part of the systematic shear correlation functions are shown in Fig. \ref{plot_ecorrIM}. The $\xi_{\rm sys}^{{\rm t}\times/\times {\rm t}}(\theta)$ measurements do not show a clear trend. The incremental weighted mean of $\xi_{\rm sys}^{\times t}(\theta)$ is consistent with zero for the three grids. $\langle \xi_{\rm sys}^{\times t}(<\theta)\rangle$, however, prefers a negative value for G2 and G3 when averaged over all scales at the 3$\sigma$ level, but the actual values are very small ($\sim-5\times10^{-8}$). Hence these correlation functions also indicate that systematics may be present in the data. \\
\indent The systematic correlation functions are in line with our results from the previous section and point at the presence of a position-dependent $c$ that is unaccounted for in the public CFHTLenS catalogues. In the next section we repeat our test on the CFHTLenS fields that did not pass the systematic tests of \citet{Heymans12}. In Sect. \ref{sec_edge}, \ref{sec_star} and \ref{sec_photoz} we investigate the source of the position-dependent $c$, and in Sect. \ref{sec_csrem} we illustrate how our correlation functions suppress the contribution from cosmic shear. It is important to stress that the amplitude of $\xi_{\rm sys}^{+}$ is much smaller than the shear correlation function. For example, \citet{Kilbinger13} measure a $\xi^+$ of $\sim5\times10^{-5}$ at a separation of a few arcmin, roughly 50 times larger than $\xi_{\rm sys}^{+}$ at those scales. Since the amplitude of the systematics is smaller than the error bars on the cosmic shear measurements when averaged over all fields, it seems highly unlikely that these systematics cause a significant bias on cosmological parameters estimates from CFHTLenS. However, as pointed out before, our test is only sensitive to an additive bias that is coherent over all pointings in the survey. An additive bias that varies between pointings would remain undetected by our method.  \\
\indent Unfortunately, correcting for a stationary position-dependent $c$ is not trivial. The averaged ellipticity in each grid cell is a combination of an intrinsic shape dispersion component and a stationary systematic ellipticity component, whose relative contributions are unknown. Fitting a functional form to the averaged ellipticities as a function of grid position does not exclusively capture the additive bias contribution, as the regions where the additive bias originates are very localised (as  $\xi_{\rm sys}^{+}$ is most discrepant from zero at small separations) but a priori unknown. Subtracting the mean ellipticity in each grid cell also removes real signal which is undesirable (but we note that this also happens, albeit to a lesser extent, in the common correction schemes for additive bias in which the average ellipticity per image is subtracted). Only if one is willing to make an assumption on the origin of the bias, for example that it is related to the PSF anisotropy, one can in principle devise a correction scheme. We recommend our tests as diagnostic tools rather than converting it into a method to correct for a position-dependent $c$. \\

\subsection{CFHTLenS fail fields \label{sec_fail}}
\begin{figure*}
  \begin{minipage}{0.4\linewidth}
      \includegraphics[width=1.\linewidth,angle=-90]{plot_CFHTLenS_whisk_ccorr_fail16.ps}
  \end{minipage}
  \hspace{10mm}
  \begin{minipage}{0.53\linewidth}
      \includegraphics[width=1.\linewidth]{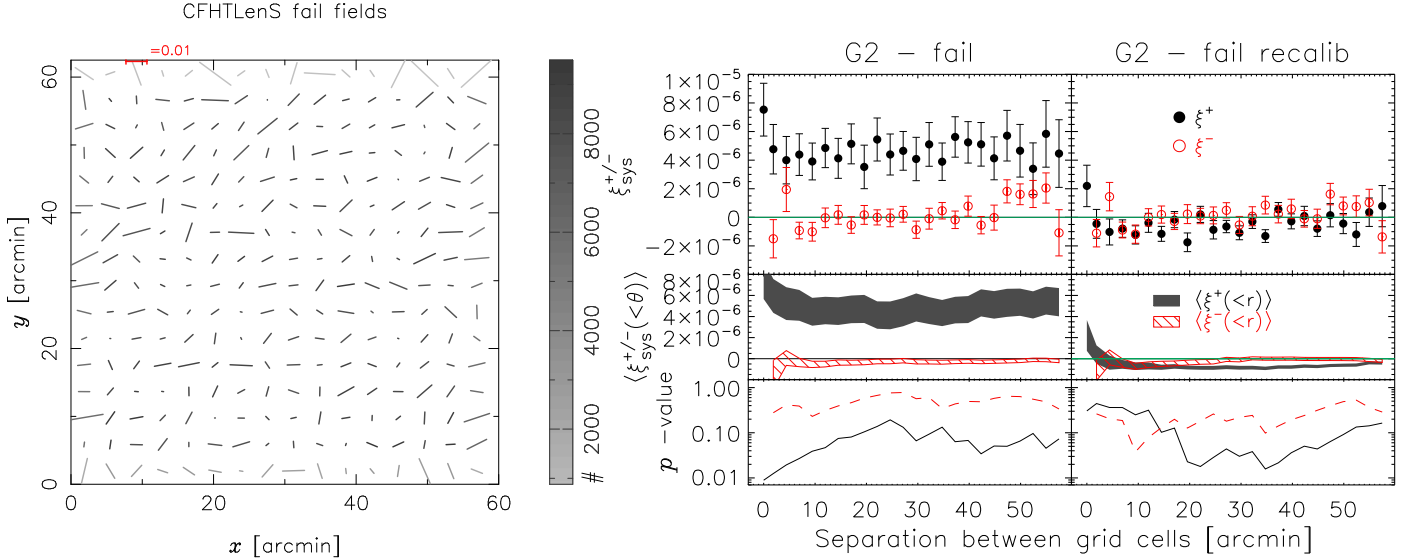}
  \end{minipage}
  \caption{{\it Left-hand panel}: Whisker plot showing the average galaxy ellipticity as a function of field position for the CFHTLenS fail fields. {\it Middle/right-hand panel}: the $\xi_{\rm sys}^{+/-}$ correlation function with and without recalibration. The middle row shows the weighted mean of $\xi_{\rm sys}^{+/-}$ and its 68\% confidence intervals, determined using all radial bins up to the one of interest. The bottom row shows the $p$-values of the null hypothesis, with the solid black (red-dashed) line for $\langle \xi_{\rm sys}^{+}(<r)\rangle$ ($\langle\xi_{\rm sys}^{-}(<r)\rangle$). The remaining trends are worse than those observed in the fields that passed the \citet{Heymans12} star-galaxy cross-correlation selection (shown in Fig. \ref{plot_ecorr}; note the different scaling of the $y$-axis).}
  \label{plot_fail}
\end{figure*}
\begin{figure*}
  \includegraphics[width=1\linewidth]{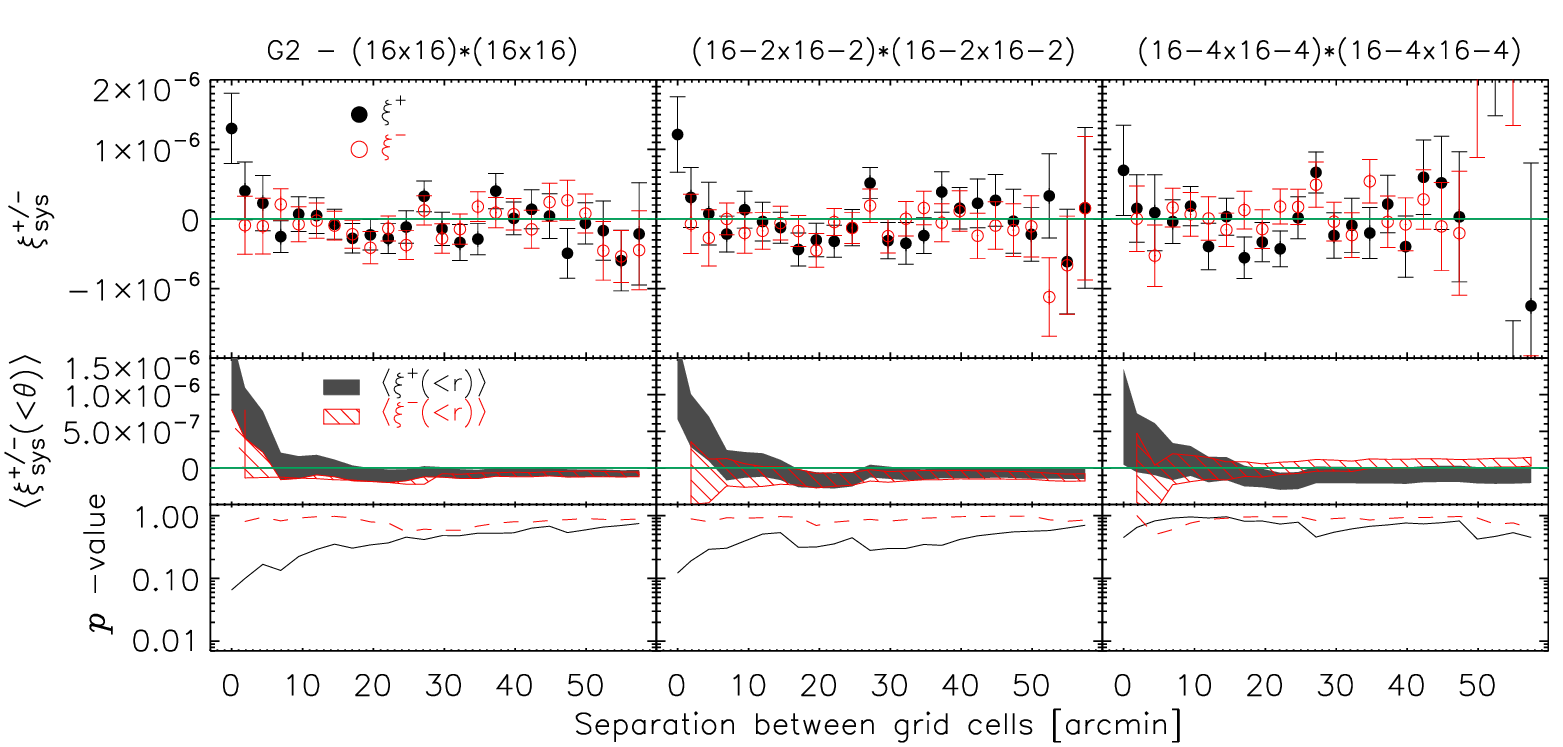}
  \caption{$\xi_{\rm sys}^{+/-}$ correlation functions as a function of separation between grid cells. In the middle panel, the signal is shown after removing one column/row at the edge of the grid, while in the right-hand panel, the first two columns/rows are removed. The left-hand panel shows the nominal signal for reference. The errors indicate the scatter between the cosmic shear-reduced bootstrap realisations. The middle row shows the weighted mean of $\xi_{\rm sys}^{+/-}$ and its 68\% confidence intervals, determined using all radial bins up to the one of interest. The bottom row shows the $p$-values of the null hypothesis, with the solid black (red-dashed) line for $\langle \xi_{\rm sys}^{+}(<r)\rangle$ ($\langle\xi_{\rm sys}^{-}(<r)\rangle$).}
  \label{plot_edge}
\end{figure*}
\citet{Heymans12} developed a novel methodology to identify fields with spurious PSF anisotropy contamination, enabling them to exclude those from their cosmological analyses. Their method consists of measuring the cross-correlation of the (PSF-corrected) shapes of galaxies with the shapes of stars in individual fields. The distribution of the magnitudes of these cross-correlations is compared to a model distribution that accounts for noise and chance alignments of the PSF with cosmic shear and intrinsic galaxy alignments. Problematic fields are identified as outliers from this model distribution. In total, 43 outliers were identified and removed from the cosmic shear analyses. \\
\indent We repeated our position-dependent $c$ tests on these so-called `fail' fields. In the left-hand panel of Fig. \ref{plot_fail} we show the whisker plot and in the middle panel the systematic correlation functions from Eq. (\ref{eq_xisys}). We find a very strong correlation for $\xi_{\rm sys}^{+}$ that is nearly independent of scale. This is suggestive of a constant additive bias. Note that neighbouring radial bins are highly correlated. \\
\indent The whisker plot suggests the presence of an overall $c_2$-term. We determined the average ellipticities of all galaxies in these fields, which are $\langle E_1 \rangle=0.0016\pm0.0003$ and  $\langle E_2 \rangle=0.0016\pm0.0003$, and recalibrated these fields by subtracting $\langle E_{1,2}\rangle$ from the galaxy ellipticities in the catalogue. The resulting systematic correlation functions are shown in the right-hand panel of Fig. \ref{plot_fail}. This correction removes the constant, scale-independent part of $\xi_{\rm sys}^{+}$. However, the correlation function still deviates from zero with a magnitude that is worse than for the CFHTLenS `pass' fields (shown in Fig. \ref{plot_ecorr}; the $y$-axis has a different scaling). Furthermore, we note that the bootstrap errors of $\xi_{\rm sys}^{+/-}$ become noticeably smaller after the recalibration. 

\subsection{Edge removal \label{sec_edge}}
\begin{figure}
  \includegraphics[width=1.\linewidth]{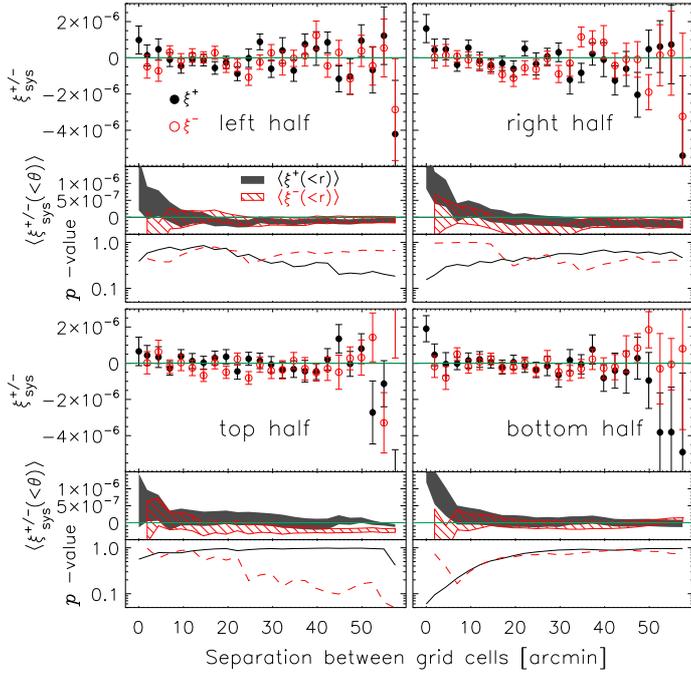}
  \caption{$\xi_{\rm sys}^{+/-}$ correlation functions as a function of separation between grid cells, computed using only half the field. The half that is used is indicated in each panel. The errors indicate the scatter between the cosmic shear-reduced bootstrap realisations. The middle row shows the weighted mean of $\xi_{\rm sys}^{+/-}$ and its 68\% confidence intervals, determined using all radial bins up to the one of interest. The bottom row shows the $p$-values of the null hypothesis, with the solid black (red-dashed) line for $\langle \xi_{\rm sys}^{+}(<r)\rangle$ ($\langle\xi_{\rm sys}^{-}(<r)\rangle$).}
  \label{plot_half}
\end{figure}

To investigate whether the position-dependent $c$ originates from a particular part of the image, we performed two tests. In Fig. \ref{plot_whisker} we found that some of the suspicious looking grid cells are located at the edge of the grid. Hence we removed the columns and rows near the edge of the grid and measured $\xi_{\rm sys}^{+/-}$ with the remaining grid cells. In Fig. \ref{plot_edge}, we show the signal after removing either 1 or 2 columns/rows near the edge of the field, for the G2 (16$\times$16) grid. The left-hand panel shows the signal for the full grid for reference.\\
\indent Excluding the grid cells near the edge of the field does not lead to a large decrease of the small-scale signal of the $\xi_{\rm sys}^+$ correlation function. The dip at $\sim$20 arcmin, however, becomes more pronounced, especially in the case where we exclude the two rows and columns near the grid edge. This shows that this dip is somehow related to a feature in the central part of the image. $\xi_{\rm sys}^{+/-}$ becomes increasingly noisy at large separations when we remove the rows and columns at the edge, because we have fewer grid cell pairs left to compute the correlation.\\
\indent Next, we measured $\xi_{\rm sys}^{+/-}$ on the left, right, bottom or top half of the grid. This allowed us to test whether certain parts of the image contain more systematics than others. If, for example, PSF residuals are larger in one of the corners, for example because of on average larger PSF anisotropies at that location, we would expect to find a larger systematic correlation function using that part of the image only. The correlation functions are shown in Fig. \ref{plot_half}. Comparing the results of the left-hand part of the grid with the right-hand part, we find that the signals look comparable. Both sides show the small-scale correlation and the dip near $\sim$20 arcmin. Comparing the top half with the bottom half, we find that the bottom half has a somewhat larger signal at small scales.  \\
\indent The dip at $\sim$20 arcmin disappears in the bottom half and is less prominent in the top half. Hence this dip is at least partly the result of a pattern in the bottom half of the grid that is anti-correlated (i.e. oriented at a relative angle of $\sim$90 degree) with a pattern in the top-half. We also note that $\xi_{\rm sys}^{+}$ becomes negative at large scales in the bottom and top half of the field, but not in the other two halves, indicating the presence of an overall anti-correlation in residual $c$ on large scales between the left-hand side and the right-hand side of the grid.

\subsection{Star density \label{sec_star}} 
One of the main causes of additive bias is thought to be the inaccurate removal of PSF anisotropies in the shape measurement process. Problems can occur at several stages: the star catalogue can be contaminated with galaxies; the model fit to the brightness profiles of stars can be inaccurate (e.g. missing the wings); the models used to capture the spatial variations of the PSF can be inadequate. The latter can occur when the PSF varies rapidly and the number of stars is insufficient to capture its small-scale variation. This would result in correlated additive biases at small scales and hence might partly explain what we observe. \\
\indent To test this, we divided all CFHTLenS pass fields in two samples, based on the density of stars used to model the PSF. PSF modelling in CFHTLenS starts with identifying star candidates from the stacked image from their location in a size-magnitude diagram. The selection is refined using the stellar loci in the $gri$ bands. The PSF is represented as a set of pixels at the same resolution of the data. The pixel values are modelled in every exposure separately, by fitting a third-order polynomial plus an additional parameter per chip, which allows for discontinuities between chips. For more details on the PSF modelling we refer to \citet{Miller13}. \\
\begin{figure}
    \centering
  \includegraphics[width=0.95\linewidth]{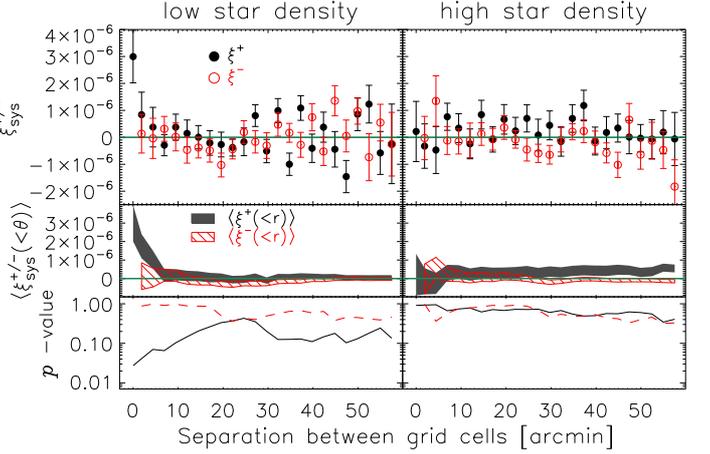}
  \caption{$\xi_{\rm sys}^{+/-}$ correlation functions, for fields with a lower than average star density (left-hand panel) and with a higher than average star density (right-hand panel). The errors indicate the scatter between the cosmic shear-reduced bootstrap realisations. The middle row shows the weighted mean of $\xi_{\rm sys}^{+/-}$ and its 68\% confidence intervals, determined using all radial bins up to the one of interest. The bottom row shows the $p$-values of the null hypothesis, with the solid black (red-dashed) line for $\langle \xi_{\rm sys}^{+}(<r)\rangle$ ($\langle\xi_{\rm sys}^{-}(<r)\rangle$).}
  \label{plot_steldens}
\end{figure}
\indent Due to seeing variations between exposures and the dithering pattern, the number of stars used to model the PSF changes somewhat between the different exposures of the same field. The variation between exposures is typically smaller than the variation between different fields. Hence we determined the average number of stars per field by counting the stars from all exposures and dividing that by the number of exposures and used that as a proxy of star density in the field. We split the CFHTLenS pass fields in a low-stellar density and a high-stellar density sample. The number of stars used in the low-stellar density sample ranges from $\sim$2300 to $\sim$3100, with a mean of $\sim$2830 stars per image. The average effective area of these images is 0.81 deg$^2$, hence this corresponds to a PSF star density of 0.97 arcmin$^{-2}$. For the high-stellar density sample, the number of stars ranges up to $\sim$9600, and the mean number of stars is $\sim$5910 per image. The average effective area of these images is 0.72 deg$^2$, hence the corresponding PSF star density is 2.28 arcmin$^{-2}$. The effective area is likely lower in the high-stellar density sample as more stars are saturated and causing reflections, which have been masked. \\
\indent We repeated the systematic tests on the two samples. The resulting systematic correlation functions are shown in Fig. \ref{plot_steldens}. For the high-stellar density fields, $\xi_{\rm sys}^{+/-}$ is consistent with zero on all scales. For the low-stellar density fields, the systematics are enhanced. This strongly suggests that the small-scale additive bias is caused by undersampling of the spatial variation of the PSF. Since the systematics have a typical scale-length of the width of a chip ($\sim$7.5 arcmin) and not that of the average separation between stars ($\sim$1 arcmin) we suspect that the parameter fitted to each chip separately in the PSF model is causing the trouble. \\
\begin{figure}
  \centering
  \includegraphics[width=1\linewidth,angle=-90]{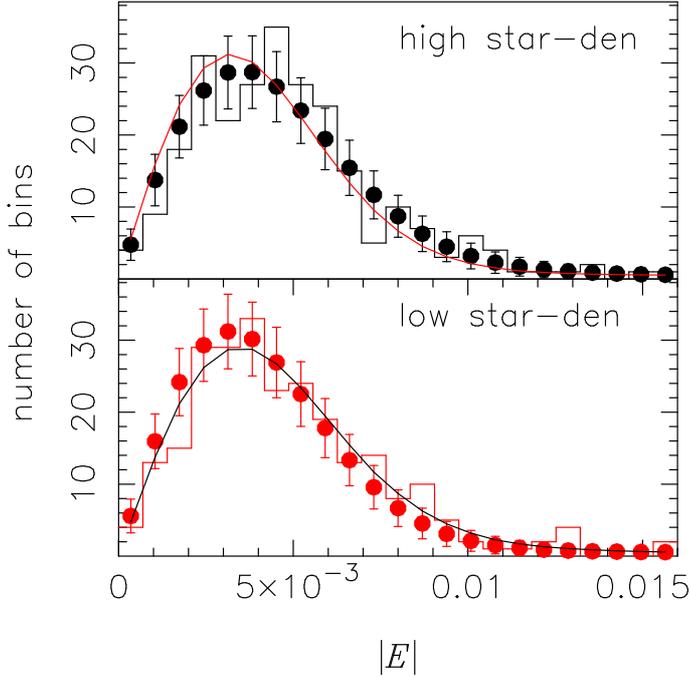}
  \caption{Histogram of mean ellipticity values in the G2 grid for the samples with higher than average ({\it top}) and lower than average ({\it bottom}) stellar density. The solid histogram shows the distribution for the original CFHTLenS catalogue, whilst the dots indicate the mean and scatter of $10\,000$ realisations where the ellipticities were randomised. The solid red line in the top panel indicates the mean of the randomised histograms of the bottom panel, and the black line in the bottom panel shows the  mean of the randomised histograms of the top panel.}
  \label{plot_ellhiststar}
\end{figure} 
\begin{figure}
  \centering
  \includegraphics[width=0.95\linewidth]{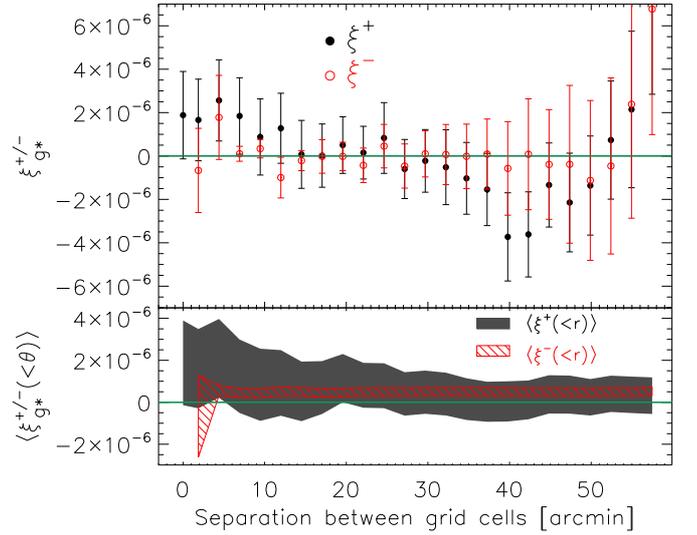}
  \caption{$\xi_{\rm g*}^{+/-}$ correlation functions as a function of separation between grid cells. The errors indicate the scatter between the bootstrap realisations. The bottom row shows the weighted mean of $\xi_{\rm g*}^{+/-}$ and its 68\% confidence intervals, determined using all radial bins up to the one of interest.}
  \label{plot_starcross}
\end{figure} 
\indent Motivated by this difference, we repeated the 1-point statistic tests on the two samples. The whisker plots are not particularly revealing and hence not shown. We do show the histograms of the ellipticity values of the whisker plots in Fig. \ref{plot_ellhiststar}. For the low-stellar-density sample, there are several bins at $|E|>6\times10^{-3}$ higher than the mean of the randomised histograms; this it not the case for the high-stellar-density sample. The reduced $\chi^2$ between the observed and the mean of the randomised histograms is 1.64 and 0.94 for the low- and high-stellar-density sample, respectively, with corresponding $p$-values of 0.028 and 0.537. The KS statistic of the two samples are 0.085 and 0.057, respectively, with corresponding $p$-values of 0.048 and 0.368. In Fig. \ref{plot_ellhiststar} we also show the mean of the randomised histogram of the lower-stellar density sample in the higher-stellar density panel, and vice versa. This shows that the dispersion of the average galaxy ellipticities in the lower-stellar-density sample is smaller than in the higher-stellar-density sample, suggesting that errors in the PSF model tend to make galaxies rounder on average. \\
\indent We also computed the systematic correlation function of the average galaxy ellipticities and the averaged PSF model ellipticities, $\langle E^{\rm *,m}\rangle$, of the same galaxies (as shown in the right-hand panel of Fig. \ref{plot_whisker}):
\begin{equation}
\begin{aligned}
& \xi_{\rm g*}^{+/-} (\theta)= \\
& \frac{\sum_{i} \sum_{j} {\sf w}_i {\sf w}_j \left ( \langle E_{\rm t}\rangle ({\bf x}_i) \langle E^{\rm *,m}_{\rm t}\rangle ({\bf x}_j) \pm  \langle E_\times\rangle ({\bf x}_i) \langle E^{\rm *,m}_\times\rangle ({\bf x}_j)\right ) }{\sum_{i} \sum_{j} {\sf w}_i {\sf w}_j},
\end{aligned}
\end{equation}
with $E^{\rm *,m}_{{\rm t}/\times}$ the tangential and cross component of $E^{\rm *,m}$. Since $\langle E^{\rm *,m}\rangle$ is not subject to cosmic shear, we can create bootstrap realisations by randomly drawing 128 fields from the full sample with replacement. The result is shown in Fig. \ref{plot_starcross}. At separations below 10 arcmin, the average galaxy ellipticity and the average PSF model ellipticity are correlated, whilst at separations of 40 arcmin they are anti-correlated. Neighbouring radial bins are highly correlated, however, which is reflected in the bottom row of Fig. \ref{plot_starcross}: when including the covariance, $\langle \xi_{\rm g*}^{+/-} (<\theta)\rangle$ is consistent with zero for all values of $\theta$.  \\
\indent Finally, we compared the star density distribution of the CFHTLenS pass and fail fields. Both distributions are similar. The fail fields do not have a spuriously low star density, and undersampling of the PSF model is not likely to be the cause of the systematics in these fields.

\subsection{Photometric redshift \label{sec_photoz}}
Most cosmic shear analyses are performed in tomographic bins, that is in narrow bins of (photometric) redshift. The average size and brightness of galaxies change with redshift. Residual PSF systematics may have different magnitudes for different populations of galaxies, and could therefore be a function of redshift as well. To test this, we measured the position-dependent $c$ as a function of redshift. \\
\begin{figure}
  \includegraphics[width=1.\linewidth]{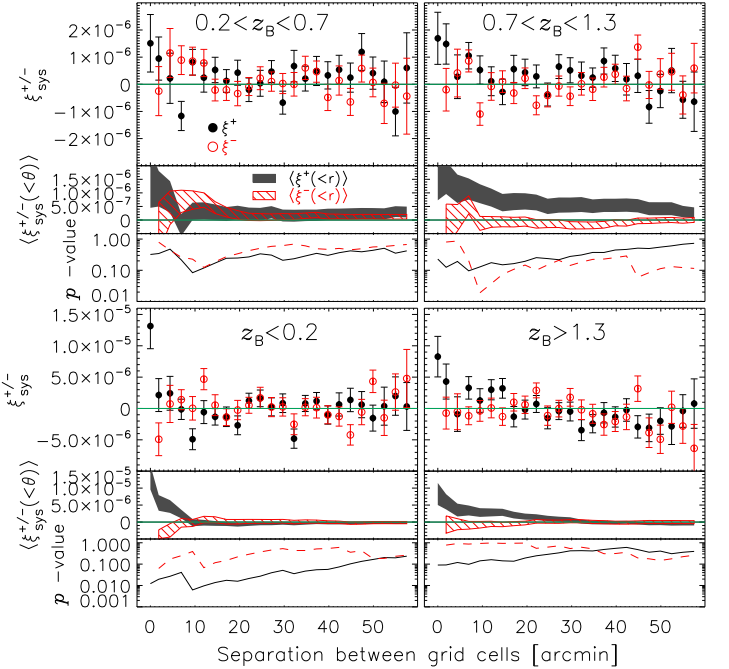}
  \caption{$\xi_{\rm sys}^{+/-}$ correlation functions as a function of separation between grid cells, computed for galaxies in different tomographic redshift bins. The top panels show the result for two tomographic bins with redshifts $0.2<z_{\rm B}<0.7$ and $0.7<z_{\rm B}<1.3$, hence covering the range where the photometric redshifts are reliable. The bottom panels show the results for galaxies outside this range. Note the different scaling on the vertical axis. The errors indicate the scatter between the cosmic shear-reduced bootstrap realisations. The middle row shows the weighted mean of $\xi_{\rm sys}^{+/-}$ and its 68\% confidence intervals, determined using all radial bins up to the one of interest. The bottom row shows the $p$-values of the null hypothesis, with the solid black (red-dashed) line for $\langle \xi_{\rm sys}^{+}(<r)\rangle$ ($\langle\xi_{\rm sys}^{-}(<r)\rangle$).}
  \label{plot_ZB}
\end{figure}
\indent In CFHTLenS, photometric redshift of galaxies are considered reliable for galaxies with $0.2<z_{\rm B}<1.3$. We split this range into two parts, $0.2<z_{\rm B}<0.7$ and $0.7<z_{\rm B}<1.3$ and show the systematic correlation functions in the top panel of Fig. \ref{plot_ZB}. $\xi_{\rm sys}^{+}$ is consistent with zero for the $0.2<z_{\rm B}<0.7$ sample on all scales, but the systematic correlation function of the $0.7<z_{\rm B}<1.3$ sample is predominantly positive (with $\langle \xi_{\rm sys}^{+}(<\theta)\rangle$ larger than zero at the 2-2.5$\sigma$ level at $\theta<40$ arcmin). However, given the larger errors, the results of $0.2<z_{\rm B}<0.7$ and $0.7<z_{\rm B}<1.3$ are consistent. Since $\xi_{\rm sys}^{+}$ of the full galaxy sample is only positive on small scales (see Fig. \ref{plot_ecorr}), but appears to be positive on most scales for the two redshift subsamples, we suspect that the systematics affecting galaxies at low and high redshift have a different pattern, such that they more or less average out for the full sample. $\xi_{\rm sys}^{-}$ is consistent with zero for most radial bins, except for a few bins around $\sim$10 arcmin for the $0.2<z_{\rm B}<0.7$ subsample, where  $\langle \xi_{\rm sys}^{-}(<\theta)\rangle$ reaches a $\sim$2.8$\sigma$ deviation from zero.\\
\indent We also measured the $\xi_{\rm sys}^{+/-}$ correlation functions in the range where the photometric redshifts are not reliable, $z_{\rm B}<0.2$ and $z_{\rm B}>1.3$. We show the results in the lower panels of Fig. \ref{plot_ZB}. The range of the vertical axis has been increased. For $z_{\rm B}<0.2$, we find a highly significant $\xi_{\rm sys}^{+}$ at zero-lag. For $z_{\rm B}>1.3$, there is also a significant $\xi_{\rm sys}^{+}$ signal at small separations, with a magnitude that is larger than for the galaxies in the reliable photometric redshift range. Additionally, there is a hint for a negative correlation at separations larger than half a degree. \\
\indent Although the galaxies with $z_{\rm B}<0.2$ and $z_{\rm B}>1.3$ are excluded in all CFHTLenS analyses, we note that the additive and multiplicative shear calibration schemes do not depend on photometric redshift. Unless the photometric redshifts in these ranges are completely bogus, the photometric redshifts are correlated to observable galaxy properties (galaxies with $z_{\rm B}<0.2$ being larger and brighter on average, whilst those at $z_{\rm B}>1.3$ smaller and fainter). This could mean that the position-dependent $c$ depends on observed galaxy properties. However, we were pointed to the fact that the model PSF ellipticity was highly correlated with the photometric redshifts if the photometry was not regaussianised. This motivated the use of regaussianised photometry in CFHTLenS. Hence it is possible that some residual correlation remains and that photometric errors in some bands places galaxies at certain redshifts (e.g. at $z_{\rm B}<0.2$). The systematics for these two redshift bins might therefore be related to PSF modelling issues and not to galaxy properties (L. Miller, {\it private communication}). Such a correlation between PSF properties and $z_{\rm B}$ was also listed in \citet{Asgari16} as a potential cause of the B-modes found in their analysis of CFHTLenS data. We postpone a further investigation of the dependence of position-dependent $c$ on galaxy properties to future work.

\subsection{Impact of cosmic shear \label{sec_csrem}}

\begin{figure}
  \includegraphics[width=1.\linewidth]{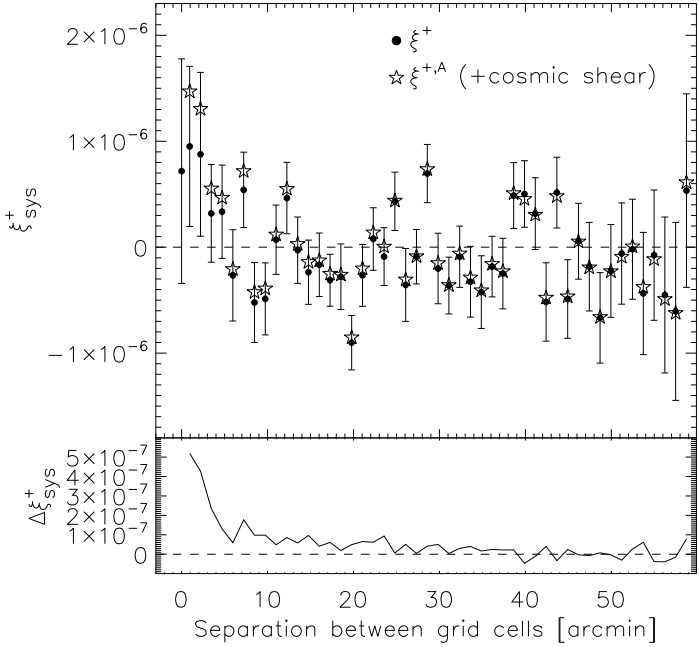}
  \caption{ $\xi_{\rm sys}^{+}$ correlation function as a function of separation between grid cells. The filled circles and the errors are the reference signal, whilst the open stars show the signal which includes a contribution from cosmic shear. The lower inset shows the difference between the two. The contribution from cosmic shear decreases with radius as expected.}
  \label{plot_errorcomp}
\end{figure}
Our method of correlating the average ellipticities of one image with the average ellipticities of the galaxies of all other images suppresses the contribution from cosmic shear. To illustrate how much cosmic shear can contribute, we correlated the average ellipticities of all images, $\langle E \rangle$, with itself:
\begin{equation}
\begin{aligned}
\xi_{\rm sys}^{+, {\rm A}} & (\theta)= \\
& \frac{\sum_{i} \sum_{j} {\sf w}_i {\sf w}_j \left [ \langle E_{\rm t}\rangle ({\bf x}_i) \langle E_{\rm t}\rangle ({\bf x}_j) +  \langle E_\times\rangle ({\bf x}_i) \langle E_\times\rangle ({\bf x}_j)\right ] }{\sum_{i} \sum_{j} {\sf w}_i {\sf w}_j}
\end{aligned}
\end{equation}
with $E_{{\rm t},\times}$ the tangential and cross components of $E$, respectively. We show $\xi_{\rm sys}^{+, {\rm A}}$ together with the reference $\xi_{\rm sys}^{+}$ in Fig. \ref{plot_errorcomp}. At large separations, the two correlation functions have similar signals, but at small scales, $\xi_{\rm sys}^{+, {\rm A}}$ is systematically larger than $\xi_{\rm sys}^{+}$, as is clear from the lower inset of Fig. \ref{plot_errorcomp} which shows $\Delta \xi_{\rm sys}^{+}=\xi_{\rm sys}^{+, {\rm A}}-\xi_{\rm sys}^{+}$. The difference is caused by cosmic shear, which causes an additional correlation between the averaged ellipticities, as neighbouring grid cells contain galaxies from the same image that are subject to the same cosmic shear field. The zero-lag point of $\xi_{\rm sys}^{+, {\rm A}}$ has a value of $\sim$$4.5\times 10^{-5}$, much higher than the rest. Since it includes the auto-correlation of galaxy ellipticities, this number is not meaningful. \\
\indent Even though our estimator suppresses the contribution from cosmic shear, there may still be some signal left from modes that stretch over several degrees. In principle, these could be further suppressed by excluding the neighbouring fields when computing $\langle E ^{\rm notS} \rangle ({\bf x}_i)$. However, the contribution should be small (much smaller than the difference between $\xi_{\rm sys}^{+, {\rm A}}$ and $\xi_{\rm sys}^{+}$), so we consider it safe to ignore it here.

\section{KiDS \label{sec_kids}}
\begin{figure}
   \includegraphics[width=\linewidth,angle=-90]{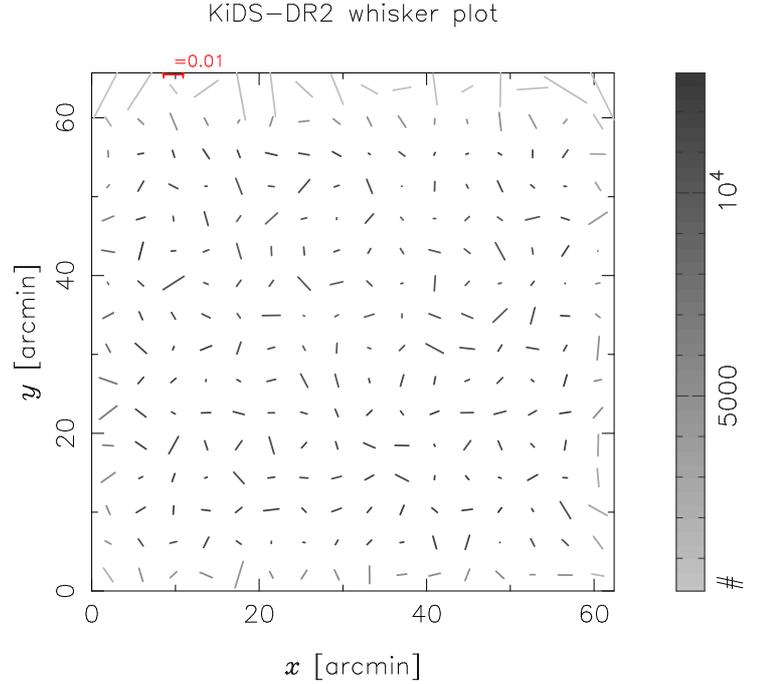}
   \caption{Average galaxy ellipticity whisker for the G2 grid (16$\times$16) for KiDS. The sticks indicate the size and orientation of the averaged ellipticities. The grey-scale of the sticks indicate the number of galaxies in a grid cell. The range of the horizontal and vertical axis corresponds to the size of a KiDS image.}
   \label{plot_whiskerK}
\end{figure}
\begin{figure*}
  \includegraphics[width=1.\linewidth]{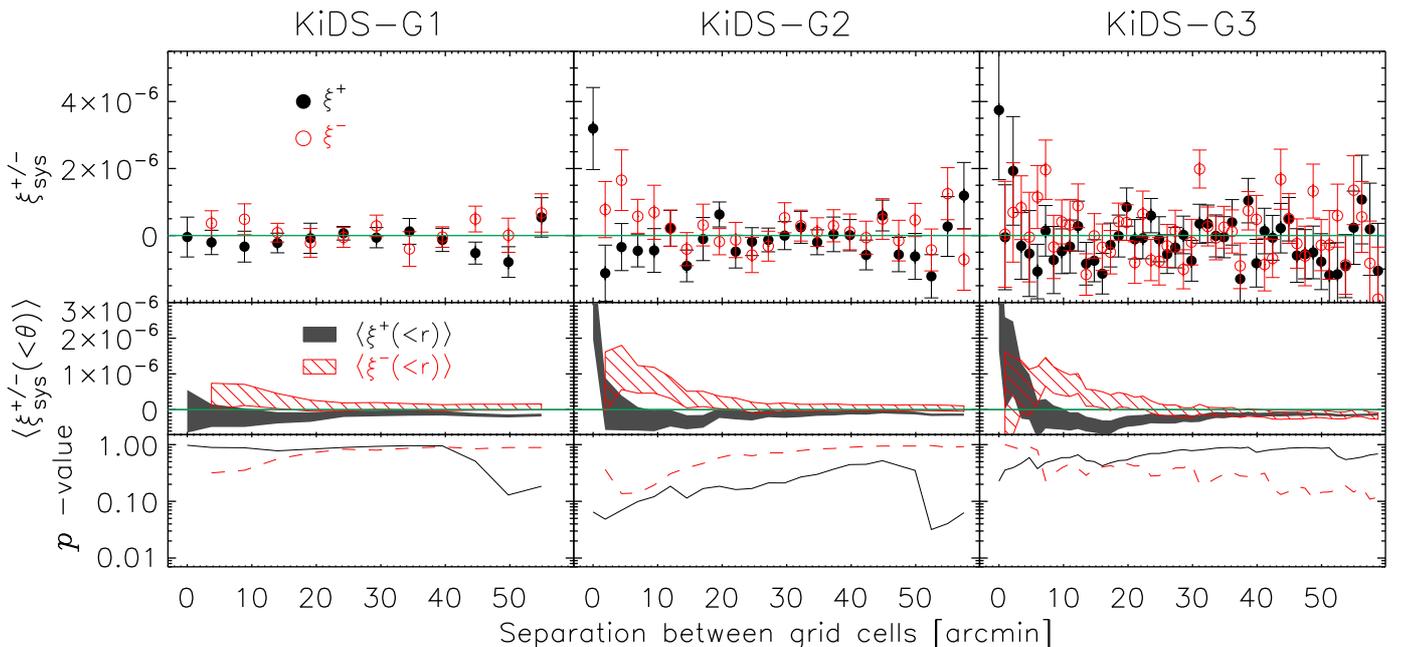}
  \caption{$\xi_{\rm sys}^{+/-}$ correlation functions as a function of separation between grid cells for KiDS. The left-hand, middle and right-hand panel show the signal when the field is split in 8$\times$8, 16$\times$16 and 32$\times$32 bins, respectively. The x-axis range corresponds to the size of the image in all three panels. The errors indicate the scatter between the cosmic shear-reduced bootstrap realisations. The middle row shows the weighted mean of $\xi_{\rm sys}^{+/-}$ and its 68\% confidence intervals, determined using all radial bins up to the one of interest. The bottom row shows the $p$-values of the null hypothesis, with the solid black (red-dashed) line for $\langle \xi_{\rm sys}^{+}(<r)\rangle$ ($\langle\xi_{\rm sys}^{-}(<r)\rangle$).}
  \label{plot_ecorrKiDS}
\end{figure*}
\begin{figure}
  \includegraphics[width=1.\linewidth]{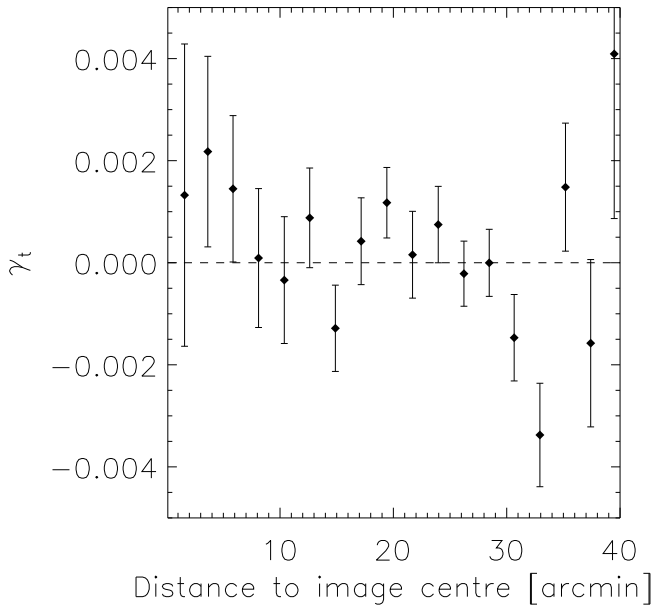}
  \caption{Average tangential shear pattern around the centre of the KiDS-DR2 images.}
   \label{plot_KiDSgt}
\end{figure}

We repeated our systematic tests on data from the Kilo Degree Survey \citep[KiDS;][]{DeJong13}. KiDS is an ongoing lensing survey that will eventually cover 1500 deg$^2$ in the $ugri$-bands. KiDS is observed with the VLT survey telescope (VST) using the OmegaCAM imager, a 1 deg$^2$ CCD camera that consists of 8$\times$4 CCDs. Each chip has 2048$\times$4096 pixels and the pixel scale is 0.21 arcsec. 109 KiDS tiles have been released as part of the first and second data release to ESO and made publicly available. The effective area after removing the data that is masked and in the overlap between tiles is 75.4 deg$^2$. \\
\indent The shear measurement procedure for KiDS is outlined in \citet{Kuijken15} and is very similar to the analysis of CFHTLenS. Galaxy shapes are measured in the $r$-band data with \emph{lens}fit and photometric redshifts are estimated from the $ugri$ photometry using BPZ; the range where the redshifts are considered reliable is \mbox{$0.005<z_{\rm B}<1.2$}. As for CFHTLenS, the additive bias is determined by averaging the ellipticities of galaxies as a function of their observed properties, but not as a function of position in the field. The additive bias is non-zero for $\epsilon_1$ and $\epsilon_2$. A strong dependence is found between $c_1$ and the Strehl ratio of the PSF, which could indicate a problem with undersampling of the PSF brightness profile \citep[not an undersampling of the spatial variation; see Sect. 5.4 in][]{Kuijken15}. The additive bias is characterised by binning the galaxies in signal-to-noise, size and Strehl ratio, to which a 3D second-order polynomial is fit. The correction factors are computed on a galaxy-by-galaxy basis and are provided as separate columns in the catalogue. \\
\indent We repeated our systematic tests on these catalogues. We applied the additive bias correction and only used galaxies from unmasked areas, with a non-zero \emph{lens}fit weight and with \mbox{$0.005<z_{\rm B}<1.2$}. We used the same grids as before. The whisker plot for G2 is shown in Fig. \ref{plot_whiskerK} and the  $\xi_{\rm sys}^{+/-}$ measurements in Fig. \ref{plot_ecorrKiDS}. $\xi_{\rm sys}^{-}$ appears to be systematically larger than zero around scales of $\sim$10 arcmin. $\xi_{\rm sys}^{+}$ is positive at zero lag, followed by a negative dip around $\sim$20 arcmin. Furthermore, we find a negative dip at scales $\sim$50 arcmin, which has to originate from regions close to the boundaries of the images. We therefore also measured the tangential shear signal around the image centres and show it in Fig. \ref{plot_KiDSgt}. It clearly shows a negative dip at separations of half a degree. The systematic signal is more clearly visible in the tangential shear than in the $\xi_{\rm sys}^{+/-}$ measurements. The tangential shear is optimised to detect tangential trends, so any systematic that is roughly tangential with respect to the image centre is more easily detected. It shows that both tests need to be done. \\
\indent Another large ongoing lensing survey is the Dark Energy Survey \citep[DES;][]{Diehl14}, whichs aims to observe 5000 deg$^2$ in the $griz$-bands. DES recently released their shear catalogues of their science verification data \citep{Jarvis15}, which covers 139 deg$^2$ up to full depth. Unfortunately, neither the pixel positions of the galaxies, nor a field-identifier were included in their catalogues, meaning that we could not repeat our systematic tests on their data. \footnote{Note, however, that whisker plots for the DES-SV catalogues are presented in Fig. 14 of \citet{Jarvis15} and seem to suggest the presence of an overall $\epsilon_2$-trend and correlated structures at small scales.}

\section{Conclusions \label{sec_concl}}
Various potential sources of additive shear bias in weak lensing data sets are a function of pixel position, such as a varying PSF anisotropy. Consequently, the additive bias itself can depend on pixel position too. Previously applied correction schemes for additive bias used field-averaged correction factors, ignoring any spatial variation that is stationary between exposures. We have developed new tests for identifying such stationary position-dependent additive shear biases. Our main test consists of determining the average galaxy ellipticities on a grid for a single image, and correlating that with the average, gridded ellipticities of the galaxies from all other fields. This test is designed to suppress the contribution from cosmic shear and hence to enhance potential remaining systematics.  \\
\indent We have applied our method to the publicly available CFHTLenS shear catalogues. After correcting the catalogues with the field-averaged additive bias corrections from \citet{Heymans12}, we found that the resulting shear whisker plot revealed some suspicious features. We first analysed the distribution of the absolute values of the ellipticities by comparing it with a randomised version of the catalogue and found that the two were significantly different. \\
\indent We quantified the positional dependence of the additive bias by measuring the systematic correlation functions, $\xi_{\rm sys}^{+/-}$ (based on Eq. \ref{eq_xisys}). $\xi_{\rm sys}^{+}$ is not consistent with zero on all scales and shows a number of features. At small scales, we find that it is positive, whilst at larger scales, there is a negative dip around $\sim$20 arcmin. The level of spurious signal is much smaller than the cosmological signal on all scales and is not expected to significantly bias the published cosmic shear results from CFHTLenS.  \\
\indent To investigate the origin of the position-dependent additive shear bias, we studied its dependence on regions in the field, on the density of stars used in the PSF modelling and on photometric redshift. We found a strong correlation with stellar density: the fields with higher than average stellar density have a systematic correlation function that is consistent with zero on all scales. This suggests that for the fields with lower than average stellar density, the number of stars used to model the spatial variation of the PSF is too low. \\
\indent Our systematic test can be trivially extended along various routes and could include trends with observed galaxy properties such as galaxy size and brightness. Another interesting option is to combine fields with similar PSF properties, such as those with a similar average seeing or with similar PSF patterns. \\
\indent Future lensing surveys will contain hundreds to thousands times more galaxies than currently used in state-of-the-art surveys like CFHTLenS. The increase of statistical power of the lensing signal needs to be matched with new methods to detect and remove ever smaller systematics in the catalogues. Identifying the positional dependence of additive shear bias is an essential one, as these can mimic and hence bias cosmic shear measurements.

\paragraph{Acknowledgements.}
We would like to thank Ludovic van Waerbeke and Lance Miller for providing the lists of stars used in the PSF modelling, and Lance Miller in addition for providing valuable feedback on this draft. EvU acknowledges support from a grant from the German Space Agency DLR and from an STFC Ernest Rutherford Research Grant, grant reference ST/L00285X/1. This work is supported by the Deutsche Forschungsgemeinschaft in the framework of the TR33 `The Dark Universe'. This work is based on observations obtained with MegaPrime/MegaCam, a joint project of CFHT and CEA/DAPNIA, at the Canada-France-Hawaii Telescope (CFHT) which is operated by the National Research Council (NRC) of Canada, the Institut National des Sciences de l'Univers of the Centre National de la Recherche Scientifique (CNRS) of France, and the University of Hawaii. This research used the facilities of the Canadian Astronomy Data Centre operated by the National Research Council of Canada with the support of the Canadian Space Agency. CFHTLenS data processing was made possible thanks to significant  computing support from the NSERC Research Tools and Instruments grant programme. This work is based on data products from observations made with ESO Telescopes at the La Silla Paranal Observatory under programme IDs 177.A-3016, 177.A-3017 and 177.A-3018. 

\bibliographystyle{aa}

\begin{appendix}

\section{Estimating $p$-values \label{app_p}}

\begin{figure*}
  \centering
  \includegraphics[width=0.7\linewidth,angle=-90]{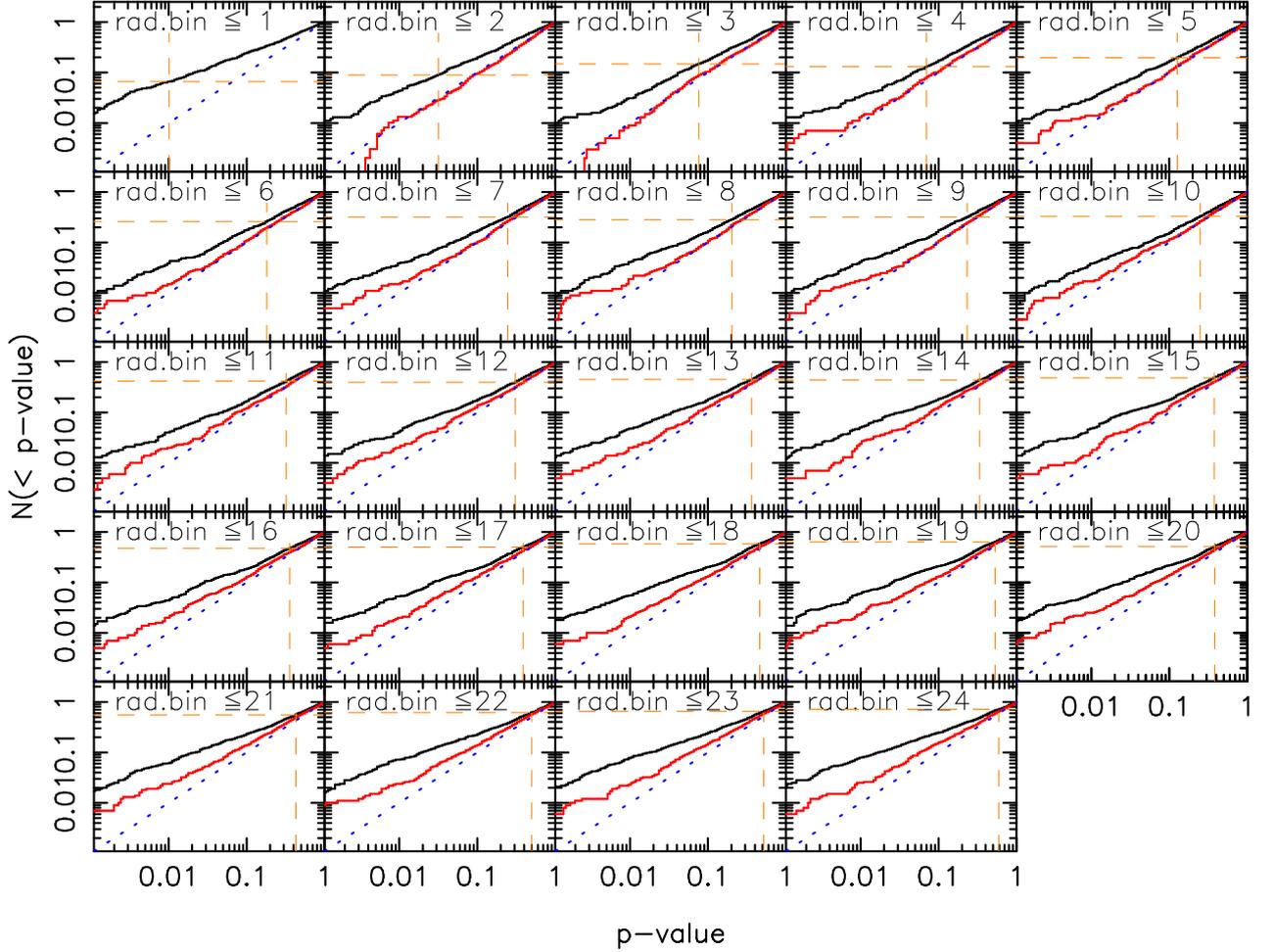}
  \caption{Cumulative distribution of the occurence of $p$-values corresponding to the $\chi^2$ value of the null hypothesis for the systematic shear correlation functions, determined using a large set of randomised catalogues. The different panels correspond, from left to right and top to bottom, to the increasing number of radial bins included in the fit (indicated in each panel). The black line shows the $p$-values of $\xi^{+}_{\rm sys}$, the red line the $p$-values of $\xi^{-}_{\rm sys}$. The blue dotted line shows the one-to-one correspondence; any departure from this line shows that the $p$-values from the $\chi^2$ of the null hypothesis do not correspond to the actual probability of $\xi^{+/-}_{\rm sys}$ being zero. The orange dashed lines show how we convert the observed $p$-value in our data to the actual probability that $\xi^{+/-}_{\rm sys}$ is consistent with zero.}
\label{plot_pvaltst}
\end{figure*}

To estimate the probability that the systematic correlation functions are consistent with zero, we computed the $\chi^2$ value of the null hypothesis, taking into account the correlation between the radial bins. The $p$-value corresponding to this $\chi^2$ value does not exactly correspond to the probability that $\xi^{+/-}_{\rm sys}$ is zero. At small scales, the number of radial bins is small and the probability derived from the $\chi^2$ is inaccurate. In addition, the errors on $\xi^{+/-}_{\rm sys}$ may not follow Gaussian statistics, which would also lead to differences between the $p$-values and the actual probabilities. \\
\indent To analyse how the $p$-values relate to the actual probabilities of $\xi^{+/-}_{\rm sys}$ being zero, we created random realisations of the data. We rotated the ellipticities of all galaxies by a random amount (different from galaxy to galaxy). Next, we analysed this randomised catalogue exactly as the real data: we measured the average ellipticities on the 16$\times$16 grid, correlated the average ellipticities of one image to the average of all the others, and estimated the covariances of the correlation functions using bootstrapping. The expectation value of $\xi^{+/-}_{\rm sys}$ of these randomised catalogues is zero by construction. We determined the $\chi^2$ value of the null hypothesis  for increasing radial scales using the covariance matrix estimated from the bootstraps, and computed the corresponding $p$-value. \\
\indent We repeated this procedure 1000 times. Then we determined the distribution of $p$-values for each radial bin. The results are shown in Fig. \ref{plot_pvaltst}. The horizontal axis of this figure shows the observed $p$-value, the vertical axis the number of times a $p$-value smaller than the observed one was found in the randomised catalogues (i.e. the real probability). If the measured $p$-values would correspond to the actual probability of $\xi^{+/-}_{\rm sys}$ being consistent with zero, one would expect a one-to-one correspondence. For $\xi^{+}_{\rm sys}$, the $p$-value generally overestimates the probability, particularly for low $p$-values (more random realisations have low $p$-values than you would expect). For $\xi^{-}_{\rm sys}$, the $p$-values agree fairly well with the probability if we only include small radial scales, but slightly overestimates the probability towards an increasing number of radial bins. \\
\indent We used this result to convert the $p$-values that correspond to the $\chi^2$ values of the null hypothesis in our measurement, to the actual probability that $\xi^{+/-}_{\rm sys}$ is consistent with zero. All $p$-values that are shown in this work have been converted like this. We derived separate conversions for each measurement. In particular, the conversion of the CFHTLenS fail fields, as well as the one of the low/high-stellar density fields, differs slightly from the one shown in Fig. \ref{plot_pvaltst} (although the main trends are the same). The conversion scheme for $\xi_{\rm sys}^{{\rm t}\times/\times {\rm t}}$ differs as well, such that the $p$-value that corresponds to the $\chi^2$ value of the null hypothesis, is closer to the actual probability. Finally, we note that for the CFHTLenS fail fields and for the $z_{\rm B}<0.2$ sample, we used ten times more random realisations, which we needed to ensure that the conversion at low $p$-values was robust.

\end{appendix}

\end{document}